\documentclass[aps,prl,twocolumn,superscriptaddress,amsfont,graphicx,nofootinbib,preprintnumbers]{revtex4}
\usepackage{color,graphicx,epsfig}
\usepackage{ifpdf}
\usepackage{amsmath}
\usepackage{bm}
\usepackage{color}
\usepackage[english]{babel}
\usepackage{graphicx}
\usepackage{amsfonts}
\usepackage{amssymb}
\usepackage[bookmarks=false]{hyperref}
\usepackage{enumerate}
\usepackage{bbold}

\definecolor{nicered}{rgb}{0.7,0.1,0.1}
\definecolor{nicegreen}{rgb}{0.1,0.5,0.1}
\hypersetup{colorlinks,citecolor= nicegreen,linkcolor= nicered}

\newcommand{\beq}{\begin{equation}}
\newcommand{\eeq}{\end{equation}}
\newcommand{\bea}{\begin{eqnarray}}
\newcommand{\eea}{\end{eqnarray}}

\definecolor{Red}{rgb}{1.,0.,0.}

\def\gsim{{~\raise.15em\hbox{$>$}\kern-.85em
          \lower.35em\hbox{$\sim$}~}}
\def\lsim{{~\raise.15em\hbox{$<$}\kern-.85em
          \lower.35em\hbox{$\sim$}~}}

\def\mysection#1{{{\bf #1}.~}}
\def\OMIT#1{}

\begin{document}

\def\Cincy{Department of Physics, University of Cincinnati, Cincinnati, Ohio 45221}
\def\NYU{Center for Cosmology and Particle Physics, Department of Physics, New York University, New York, NY 10003}
\def\ND{Department of Physics, 225 Nieuwland Science Hall, University of Notre Dame, Notre Dame, IN 46556}

\title{A UV complete partially composite-pNGB Higgs}

\author{Jamison Galloway}
\email[Electronic address:]{jamison.galloway@nyu.edu}
\affiliation{\NYU}

\author{Alexander L. Kagan}
\email[Electronic address:]{kaganal@ucmail.uc.edu}
\affiliation{\Cincy}

\author{Adam Martin}
\email[Electronic address:]{amarti41@nd.edu}
\affiliation{\ND}

%\date{\today}

\begin{abstract}
We explore an electroweak symmetry breaking (EWSB) scenario based on the mixture of a fundamental Higgs doublet and an SU(4)/Sp(4) composite pseudo-Nambu-Goldstone doublet -- a particular manifestation of bosonic technicolor/induced EWSB. Taking the fundamental Higgs mass parameter to be positive, EWSB is triggered by the mixing of the doublets. This setup has several attractive features and phenomenological consequences, which we highlight: i) Unlike traditional bosonic technicolor models, the hierarchy between $\Lambda_{\rm TC}$ and the electroweak scale depends on vacuum (mis)alignment and can be sizable, yielding an attractive framework for natural EWSB; ii) As the strong sector is based on SU(4)/Sp(4), a fundamental (UV-complete) description of the strong sector is possible, that is informed by the lattice;
iii) The lightest vector resonances occur in the 10-plet, 5-plet and singlet of Sp(4).  
Misalignment leads to a 10-plet  ``parity-doubling" cancelation in the $S$ parameter, and a suppressed 5-plet contribution; iv) Higgs coupling deviations are typically of $\mathcal O(1\%)$; v) The 10-plet isotriplet resonances decay dominantly to a massive technipion and a gauge boson, or to technipion pairs, rather than to gauge boson or fermion pairs; moreover, their couplings to fermions are small. Thus, the bounds on this setup from conventional heavy-vector-triplet searches are weak. A supersymmetric $U(1)_R$ symmetric realization is briefly described.
\end{abstract}
\maketitle
%
%%%%%%%%%%%%%%%%%%%%%%%%%%%%%%%%%%%%%%%%%%%%
%

\section{Introduction}
The minimal supersymmetric Standard Model (MSSM) 
provides an elegant mechanism for stabilizing the Higgs mass, and benefits from the simplicity of the Yukawa coupling paradigm for fermion mass generation. However, naturalness in the MSSM is challenged 
by LHC bounds on colored superpartners.   
Technicolor (TC) provides a beautiful mechanism for electroweak symmetry breaking (EWSB), based on asymptotically free gauge theories, but a light Higgs is difficult to accommodate.  
The pseudo-Nambu Goldstone boson (pNGB) composite Higgs improves on TC by providing a large gap between the Higgs mass and other strong interaction resonances, via vacuum (mis)alignment.
However, it is not obvious that a sufficiently light pNGB Higgs obtains in explicit strong interaction constructions, and a UV complete model of fermion masses is difficult to achieve, as in TC.

We introduce a promising framework for naturalness, which combines the advantages of the three approaches,
without the potential drawbacks.
It is based on Bosonic Technicolor (BTC)/induced electroweak symmetry breaking~\cite{Simmons:1988fu, Samuel:1990dq,
  Dine:1990jd, Kagan:1990az,Kagan:1991gh,Kagan:1991ng,Kagan:1992aq,Carone:1992rh, Carone:1993xc,Dobrescu:1995gz,Dobrescu:1997kt, Dobrescu:1998ci,kagansantabarbara, Antola:2009wq,Azatov:2011ht, Azatov:2011ps, Carone:2012cd,Galloway:2013dma,Chang:2014ida, kagantalks, Beauchesne:2015lva, Harnik:2016koz, Alanne:2016rpe, Gherghetta:2011na},
with $SU(2)_{\rm TC}$ gauge group and two fundamental flavors ($n_f$=2).  
In BTC, the vacuum expectation value of a fundamental Higgs (with Yukawa couplings to the Standard Model fermions) is induced from TC dynamics,
via its Yukawa couplings to the technifermions; and supersymmetry (SUSY) is introduced to protect the Higgs mass  \cite{Samuel:1990dq, Dine:1990jd, Kagan:1990az,Kagan:1991gh,Kagan:1991ng,Kagan:1992aq,kagansantabarbara,
  Azatov:2011ht, Azatov:2011ps,Galloway:2013dma,Chang:2014ida,kagantalks, Gherghetta:2011na}. The TC superpartners decouple above the TC chiral symmetry breaking scale, $\Lambda_{\rm TC}$.
From the point of view of the low energy scalar potential, BTC can accommodate a wide range of Higgs masses, including $m_h = 125$ GeV,
without fine-tuned cancelations. 

The strong sector of minimal BTC has an $SU(4)/Sp(4)$ coset description in the IR, allowing for non-trivial vacuum alignment between the $SU(2)_L$ conserving (EW vacuum) and $SU(2)_L$ breaking (TC vacuum) limits, thus yielding a composite pNGB Higgs \cite{Galloway:2010bp,Gripaios:2009pe,Barnard:2013zea,Cacciapaglia:2014uja,Ferretti:2014qta,Arbey:2015exa}. 
Prior BTC studies have existed in the TC vacuum, where the would-be pNGB Higgs decouples.   
In this work, we explore small misalignment from the EW vacuum, exploiting the interplay between EW conserving and EW breaking constituent masses. The fundamental and pNGB scalars mix, yielding a partially composite-pNGB Higgs.   There are several benefits:
(i) vacuum misalignment yields a separation of scales, allowing $\Lambda_{\rm TC}$ to be raised well into the multi-TeV region;
 (ii) the Higgs mass parameter $m_H$ is also increased.  In a supersymmetric realization, raising $\Lambda_{\rm TC}$ and $m_H$ would allow the SUSY breaking scale to be raised, which is desirable for natural EWSB. 
A $U(1)_R$-symmetric example is briefly discussed in the concluding remarks;
(iii) vacuum misalignment reduces deviations from the SM in the Higgs couplings and precision EW parameters, the latter 
due in large measure to $SU(2)_{L\leftrightarrow R}$ parity doubling in the vector sector.

\section{The UV theory}
In this letter our focus is on the impact of the UV theory on the BTC vacuum structure, and the resulting scalar and vector meson masses and interactions.
For this purpose it suffices to 
consider minimal non-supersymmetric BTC, with a single Higgs doublet $H$. The extension to two Higgs doublets is straightforward.
The minimal TC sector contains the gauge group $SU(2)_{\rm TC}$, together with an $SU(2)_L$ doublet and two singlet technifermions, i.e. $n_f =2$, 
see Table \ref{tab:model}. 
\begin{table}[]
\begin{center}
\renewcommand{\arraystretch}{1.1}
\begin{tabular}{c | c | c c}
& $SU(2)_{\rm TC}$ & $SU(2)_W$ & $U(1)_Y$ \\
\hline
$ (\Psi^1 \,\Psi^2 )^T \equiv T_{1,2}$ & $(\square\,\,\square )^T$ & $\square$ & $0$ \\
$\Psi^3 \equiv U$& $\square$ & $1$ & $-1/2$ \\
$\Psi^4 \equiv D $& $\square$ & $1$ &\hspace{0.1cm} $+1/2~$.
\end{tabular}   
\caption{\small{Technifermion gauge quantum numbers.}}
\label{tab:model}
\end{center}
\end{table} 
All of the technifermions are treated as left-handed Weyl fields, transforming under the $(1/2, 0)$ representation of the Lorentz group $SU(2) \times SU(2) \sim SO(3,1)$.  
With weak interactions turned off, the model possesses a global $SU(4)$ symmetry under which the four-component 
 object
\beq \Psi =(T_1 \,T_2 \,U \, D)^T \, \eeq
transforms as a fundamental.

The TC condensate 
\beq \langle \Psi^{a} \,\Psi^{T,b}  \epsilon \,C^{-1}  \rangle \propto \Phi^{ab}\, \eeq
is antisymmetric in the $SU(4)$ flavor indices $a,b$; the matrix $C$ is defined momentarily.
We assume that $\Phi$ breaks $SU(4) $ to its maximal vectorlike subgroup $Sp(4)$,
yielding an $SU(4)/Sp(4)$ coset structure.
The most general $Sp(4)$ preserving condensate is  \cite{Galloway:2010bp}
\beq 
\label{eq:vacuumCP}
\Phi = \begin{pmatrix}
e^{i \alpha} \,\epsilon \cos \theta &{\mathbb{1}}_2 \sin \theta \cr -{\mathbb{1}}_2 \sin \theta & -e^{-i \alpha} \epsilon \cos \theta \end{pmatrix},  \eeq
where $\theta \in [0,\pi]$ and $\alpha$ is a CP violating phase ($\Phi \mapsto - \Phi^\dagger$ under $CP$, see \eqref{CPop}).
At $\sin\theta=0$ electroweak symmetry is unbroken, while at $\sin\theta =1 $ the condensate is purely $SU(2)_L$ breaking.  
These limits are referred to as the electroweak (EW) and TC vacua, respectively. 

The $Sp(4)$ vacuum degeneracy 
is lifted by the UV technifermion interactions. 
Previous BTC studies only included the Higgs Yukawa couplings, which selects the TC vacuum, $\sin\theta =1$.
We will explore the benefits of misalignment from the electroweak vacuum,  or small to moderate $\sin\theta$.
This is minimally accomplished by adding gauge singlet technifermion masses of $\mathcal O ( v_W  )$.  (They can be linked to 
SUSY breaking and, therefore, to $\Lambda_{\rm TC}$.)
The UV potential In $SU(4)$ notation is 
\beq  V_{UV}  =- \Psi^T \,\epsilon \,C^{-1} ( M + \lambda ) \,\Psi + h.c. +m_H^2 |H|^2 + \lambda_h |H|^4  \label{LUV}  \,\,\eeq
where $C^{-1} ={\rm diag}[i \sigma_2 , i \sigma_2 , i \sigma_2, i \sigma_2] $ acts on the LH Weyl spinors in $\Psi$, $\epsilon $ acts on the 
TC indices, and $H$ is the SM Higgs doublet with $m_H^2 > 0$ and quartic coupling $\lambda_h$.  The $4\times 4$ matrices $M $ and $\lambda $ contain the gauge singlet masses $m_{1,2}$ and the Higgs Yukawa couplings, respectively,
\beq 
M =\frac12 \begin{pmatrix}
m_1 \,\epsilon & 0  \cr
 0 & -m_2 \,\epsilon
\end{pmatrix},~~~~~\lambda=\frac12 \begin{pmatrix}
0 &  -{H}_\Lambda \cr
{H}_\Lambda^T & 0 
\end{pmatrix}\label{lambdadef} \,,
\eeq
where 
\beq \!\!\! H_\Lambda = {1\over \sqrt{2}}  \begin{pmatrix}
\lambda_U ( {\sigma_h + v^*- i \pi_h^3  })  & \lambda_D (- i \pi_h^1 + \pi_h^2 ) \cr
-\lambda_U ( i \pi_h^1 + \pi_h^2 ) & \lambda_D ( {\sigma_h + v +  i \pi_h^3 })
\end{pmatrix}.\label{HLambdagen}
\eeq
$\sigma_h $ ($\vec{\pi}_h $) are the scalar (pseudoscalar) components of $H$, 
with $v\equiv | \langle H \rangle |$.
The fermion masses are
\beq m_1 T_2 T_1 + m_2  U D + m_U T_1 U + m_D T_2 D \,,\eeq
where $m_U = \lambda_U v^* /\sqrt{2}$, $m_D = \lambda_D v /\sqrt{2}$.
Under $SU(4)$ rotations, $(M+\lambda) \mapsto U^* (M+\lambda )\, U^T$ with $U \in SU(4)$.

The gauge-kinetic term for $\Psi$, including the electroweak and TC interactions is
\beq 
{\cal L}_{\rm KE} = i \Psi^\dagger \,\bar \sigma^\mu (\partial_\mu-i  {\cal A}_\mu - i  G^{a}_\mu  { \tau^a \!/2 }\,\, {\mathbb{1}_4} )\,\Psi \,,\label{ewkKE}\eeq
where $\bar \sigma_\mu = (1  , - \vec{\sigma}^i_\mu )$ and 
\beq {\cal A}_\mu = {\rm diag}[g_{2} \,W_{\mu}^a \,\frac12 \tau^a,- g_{1} B_{\mu}  \,\frac12 \tau^3].\label{ewkKEmatrix}\eeq 

We preface our analysis of the IR with a brief discussion of discrete symmetries in the UV theory.  
With the weak interactions turned off, the only discrete symmetry of the gauge-kinetic Lagrangian
lying outside of $SU(4)$ is $CP$, under which $\Psi$ transforms as
\beq CP:~\Psi (x^\mu )  \mapsto 
 i \,\epsilon \,C^{-1} \, \Psi^*(x_\mu ) \,.\label{CPop} \eeq
Due to the pseudoreality of the $SU(2)_{\rm TC}$ fundamental, $P$ and $C$ are separately unphysical, only being defined up to arbitrary 
$SU(4)$ rotations.  The EW interactions in \eqref{ewkKE} are $CP$ invariant, with ${\cal A}_\mu \mapsto {\cal A}_\mu^T $
under $CP$.
For simplicity, we take real $m_{1,2}$, $\lambda_{U,D}$, i.e. $CP$ invariant $V_{UV} $.  We have checked to $\mathcal O(p^4)$ in the chiral expansion \cite{GKMlong}
that  this yields $\alpha = {\rm arg}(v) = 0$ at the minimum of the potential (if NDA
is not grossly violated), which we assume below.

${ G}_{LR} $-parity interchanges the generators of $SU(2)_L $ and $ SU(2)_R$. 
It is an element of the unbroken $Sp(4)$,   
transforming $\Psi \mapsto 
{\cal G}_{LR} \Psi $, with
\beq {\cal G}_{LR} = -\begin{pmatrix} 0 &  \sigma_2 \cr \sigma_2 & 0 \end{pmatrix} \,, \label{GLRdef}\eeq
up to an overall phase.
This can be seen by extending to the left-right symmetric gauge group, i.e. replacing $g_1\,B_\mu \tau^3 \mapsto  g_{2R} W_{R\,\mu}^a \tau^a $, and requiring that under $G_{LR}$ the top and bottom two components of $\Psi$ are exchanged, and $g_{2L} W_L \leftrightarrow g_{2R} W_R$.  
$G_{LR}$ invariance of $V_{UV} $ would imply $m_1 = m_2 $ and $\lambda_U = \lambda_D$.

\section{The vacuum alignment and scalars}
The $SU(4)/Sp(4)$ coset contains   
five broken generators $X^a$ in the 5-plet of $Sp(4)$,
and ten unbroken ones $T^a$ in the adjoint, satisfying $ X \Phi - \Phi X^T =0$, $\, T \Phi + \Phi T^T =0$  \cite{Galloway:2010bp}.
The isotriplets are
$T^{a=1,2,3} = {\rm diag}[\tau^a, (-)^a \tau^a ]\,/2 \sqrt{2}$, 
 \beq \!\!\!\!T^{a=4,5,6} ={1\over 2\sqrt{2} } \begin{pmatrix} c_\theta  \tau^{a-3} & - i s_\theta \tau^{a-3}\,\tau^2  \cr  -i (-)^{a} s_\theta \tau^{a-3}\,\tau^2 & \,\,(-)^a\, c_\theta\ \tau^{a-3}\end{pmatrix} ,\eeq
and $X^{a=1,2,3}$, obtained via $c_\theta\to s_\theta$, $s_\theta\to-c_\theta$ in $T^{a+3}$.
The other generators are listed in \cite{Galloway:2010bp} (with $T^{7,\dots,10}$ denoted $T_\parallel^{1,\dots,4}$).
The 5-plet decomposes as $(2,2) + (1,1)$ under the $Sp(4) $ subgroup $SU(2)_1\times SU(2)_2$, where
$SU(2)_{1,\,2}$ are identified with the generators $(T^{a}  \pm T^{a+3} )/\sqrt{2}$, $a$=1,2,3, and reduce to $SU(2)_{L,\,R}$ in the $\sin\theta\to 0 $ limit. 
$T^{1,2,3}$ are the generators of the isospin group $SU(2)_{V} = SU(2)_{L+R}=SU(2)_{1 +2} $.

Following \cite{CCWZ}, the $Sp(4)\cong SO(5)$ 5-plet of Nambu-Goldstone bosons (NGB's) $\vec{\pi}$ appears in the exponential
\beq \xi =\, \exp (\sqrt 2 i \, \pi^a X^a/f)   \mapsto \, U \xi V^\dagger,\eeq
where $f$ is the TC decay constant in the chiral limit and the transformation applies to the global rotations $U \in SU(4)$, $V\in Sp(4)$, thus $V \Phi V^T = \Phi$.
The  $\pi^a$ transform under $CP$ like the vector currents $\Psi^\dagger\bar\sigma_\mu X^a \Psi  $, see \eqref{CPop}, and similarly for $G_{LR}$, yielding 
$CP$-odd (even) $\pi^{1,3,5}$ ($\pi^{2,4}$), and $G_{LR}$-odd (even) $\pi^{1,2,3,5}$ ($\pi^4$). 

The kinetic terms are expressed in terms of 
$C_\mu = i \xi^\dagger D_\mu \xi $.  Projecting onto the broken and unbroken directions 
defines ($C_\mu = d_\mu + E_\mu $)
\beq 
\begin{split}
d_\mu & = 2 {\rm tr}(C_\mu X^a ) X^a \mapsto V d_\mu V^\dagger ,\cr
E_\mu &= 2 {\rm tr}(C_\mu T^a ) T^a \mapsto V (E_\mu + \partial_\mu ) V^\dagger\,,
\end{split}
\eeq
which, respectively, are a 5-plet and 10-plet of $Sp(4)$, transforming homogeneously and like a gauge field, as indicated.
We further define the building blocks
\beq \chi_\pm = \xi^T (M+\lambda ) \xi \Phi \pm h.c. \label{chipm}\,,\eeq
transforming as $\chi_\pm \mapsto V \chi_\pm V^\dagger $ under $Sp(4)$.

The leading $\mathcal O(p^2) $ chiral Largangian is 
\beq {\cal L}^{(2)} = {f^2\over 2} {\rm tr} ( d_\mu \,d^\mu )  +  4   \pi f^3 Z_2 \,{\rm tr} ( \chi_+ ),\label{Lp2}\eeq
where $Z_2 \approx 1.47$ at this order, according to
a recent $N_c = n_f =2$ lattice study \cite{Arthur:2016dir}.
The TC and Higgs gauge-kinetic terms yield the EW scale ($v_W = 246$ GeV)
\beq v_W^2 =  f^2 \sin^2 \theta + v^2 \, .\label{vWreln} \eeq
Minimizing the $\mathcal O(p^2)$ potential ($m_{12} = m_1 + m_2$, $\lambda_{UD} = \lambda_U + \lambda_D$) \beq
\label{V02}
V_{\rm eff}^{(2)} = 8\pi f^3 Z_2 (m_{12}  \cos\theta - {\lambda_{UD} v}\sin\theta/\sqrt{2}) +m_H^2  v^2 /2 \,,
\eeq
yields ($m_{UD} , m_{12} >0$ and $\theta \in [\pi/2,\pi]$) 
\beq  
\begin{split}
\tan\theta & = - {m_{UD} \over m_{12 }} ,~~v = {4 \sqrt{2} \,\lambda_{UD} \, \sin\theta \, f^3 \pi  Z_2 \over m_H^2 } \cr
&~{\Rightarrow}~~\sin\theta = \sqrt{1 -{m_{12}^2 \over \lambda_{UD}^4 }\,{m_H^4 \over 16  \pi^2 f^6 Z_2^2 } }.\label{vacuum}
\end{split}
\eeq
For simplicity, we have ignored the quartic in \eqref{V02}, motivated by SUSY BTC where it is a small perturbation.  The effects of EW gauge boson loops, which favor a vacuum alignment $\sin \theta \to 0$ \cite{Peskin:1980gc, Preskill:1980mz}, also constitute a small perturbation provided $m_{12}/f \gg \alpha_{\rm EW}$; this is indeed the case in the numerical examples below and thus we ignore such effects in what follows.  The limit $m_{12}/f \lsim \alpha_{\rm EW}$
%(with small $\sin\theta$ corresponding to $m_{UD} = O({\rm few})$ GeV) 
is also of interest\footnote{In this case, $\tan\theta = \mathcal O(m_{UD} /\alpha_W f )$, and small to moderate $\sin\theta $ would correspond to $m_{UD} = \mathcal O({\rm few})$ GeV, or $\lambda_{UD} = O(10^{-2})$.}, and will be considered elsewhere \cite{GKMlong}.

To elucidate the structure of the vacuum and scalar mass matrices, we project $(M+\lambda)$ onto the $Sp(4)$ singlet ($\propto \Phi$ below) and vector directions, yielding 
\bea M +  \lambda &=&- {1\over 2} \Big(  \hat m + {\lambda_{UD} \sigma_h + i\, \delta \lambda_{UD} \, \pi_h^3 \over 2 \sqrt{2}} s_\theta \Big) \,\Phi ~~~~\nonumber \\
&+& {i \over 2} 
\Phi\, \left(\lambda_{UD}  \,\chi^a_\theta    + i \delta \lambda_{UD} \, \ \chi^{\,\prime a}_{\theta} \right) X^a \! , \label{MlambdaSO5}
\end{eqnarray}
where the $Sp(4)$ singlet fermion mass and vectors are 
\bea  \hat m&\equiv&{\tiny \frac12}( -  m_{12}\, c_\theta +  m_{UD}\, s_\theta) =2 \pi f^3 Z_2 \lambda_{UD}^2 /m_H^2 \,|_{_{\theta < \pi}} \nonumber\\
\vec{\chi}_\theta& =&( \pi_h^1,\,\pi_h^2,\,\pi_h^3 ,\sigma_h c_\theta  +v \,c_\theta + \sqrt{2} \,m_{12}s_\theta /\lambda_{UD} , 0) \nonumber \\
&=&( \pi_h^1,\,\pi_h^2,\,\pi_h^3 ,\sigma_h c_\theta  , 0)\nonumber \\    
\vec{\chi}^{\,\prime}_\theta &=&(-\pi_h^2 ,\, \pi_h^1 ,\, \sigma_h +v, \pi_h^3 \,c_\theta ,\, \delta m_{12} /\delta \lambda_{UD}),
\label{chidefs}\eea
$c_\theta  \equiv \cos\theta$, $\delta m_{12} \equiv m_1- m_2$, etc., and the $O(4)$ components of $\vec\chi_\theta$,
$\vec{\chi}^{\,\prime}_\theta$ have opposite $CP$ \cite{Gasser:1983yg}.
The constant terms in $\vec{\chi}_\theta$ must cancel to avoid a constant$\,\times \pi^4$ term in $V_{\rm eff}$, induced by operators $\propto \vec{\chi}_\theta \cdot \vec{\pi}\,$. 
Thus,
$t_\theta = - m_{UD}/m_{12}$ holds to all orders.
$\vec \pi_h $ and $\pi^{1,2,3}$ are aligned, being $SU(2)_V$ triplets, 
however  
$\pi^4 $ is rotated by $\theta$ relative to $\sigma_h $. 

The $\mathcal O(p^2)$ charged scalar and neutral Higgs mass matrices are
\beq 
\begin{split}
&~~~~~~M_{\pi^+}^2  =  m_H^2 \begin{pmatrix}
1 &-{ t_\beta} \cr
-{  t_\beta} &t_\beta^2 \cr
\end{pmatrix},\cr
M_h^2 =  &  m^2_H     \begin{pmatrix}
c_\theta^2  &-{ c_\theta t_\beta} \cr
-{ c_\theta t_\beta} &t_\beta^2 \cr
\end{pmatrix} +\begin{pmatrix}
 {m}_H^2 s_\theta^2 &0 \cr
0 &0
\end{pmatrix},  \label{Mhsqpisq}
\end{split}
\eeq
in the bases $(\pi_h^+ , \pi^+ )$ and $(\sigma_h, \pi^4 )$, respectively, where
\beq t_\beta \equiv \tan\beta = v/(f \sin\theta ).\label{tbeta}\eeq
$M_{\pi^+}^2$ and the first term in $M_h^2$ are related by $Sp(4)$ invariance: their (1,1), (1,2), and (2,2) entries 
are $\propto \vec{\chi}_\theta\cdot \vec{\chi}_\theta$, $\vec{\chi}_\theta \cdot \vec{\pi}$, and $\vec{\pi}\cdot \vec{\pi}$, respectively. (The (2,2) entries correspond to the Gell-Mann-Oakes-Renner pion mass relation for  
fermion mass $\hat m$, i.e.
$m_\pi^2 =  m_H^2 t_\beta^2=16 \pi f Z_2  \,\hat m$.)
Both matrices have massless eigenstates: the eaten NGB's $G^a$ and would-be light Higgs.  The latter's mass is lifted by the 
contribution of the second term in $M_h^2$ to the $Sp(4)$ singlet, 
$\propto (\sigma_h s_\theta )^2$.
The mass eigenstates are 
\beq 
\begin{split}
 G^{\pm}& = s_\beta \, \pi^\pm _h +c_\beta \, \pi^\pm ,~~~~
 \tilde \pi^\pm=-c_\beta \, \pi^\pm_h + s_\beta \,\pi^\pm ,~~\cr
h_1&= c_\alpha\, \sigma_h  - s_\alpha \, \pi^4 ,~~~~
h_2 = s_\alpha\, \sigma_h  +c_\alpha \, \pi^4 ,\end{split}\label{masseigenstates}\eeq
 where $\tan 2 \alpha =\cos\theta \,\tan 2\beta$. The non-zero masses are
\beq  m^2_{\tilde \pi} = m_H^2 / c_\beta^2,~~~
m^2_{h_{1,2}}   = m_H^2 \left(1 \mp \sqrt{1-s_\theta^2\, s_{2\beta}^2 }\right)/(2 c_\beta^2) \,.\label{scalarmasses}
\eeq
where $ h_1$ and $h_2$ are the light and heavy neutral Higgs, respectively.
There are additionally two neutral pion states, $G^0$ and $\tilde \pi^0$,
and an isosinglet, $\pi^5$, with mass $m_H^2 t_\beta^2$.
In the limit\footnote{The Higgs quartic is included in \eqref{Mhsqpisq} by substituting $m_H^2 \to m_H^2 + \lambda_h^2 v^2/2$
and, additionally, shifting $(M_h^2 )_{1,1} $ by $\lambda_h v^2$, thus perturbing $m_{h_1}^2$ by $ \approx  \lambda_h v^2$ and $\tan\alpha$ by $ \approx 
t_\alpha c_\beta^2 \lambda_h v^2 /m_H^2 $.}   $s_\theta^2 \,c_\beta^2 \ll 1$,
\beq m_{h_1}^2 =  m_H^2  \sin^2 \beta \sin^2\theta . \label{mhp2}\eeq
Thus, a dominantly fundamental Higgs with subleading composite pNGB component ($\pi^4$ is massless for $m_{1,2}, \lambda_{U,D} \to 0$) acquires its mass from 
strong sector vacuum misalignment, as in composite pNGB Higgs models, see e.g. \cite{Agashe:2004rs,contino, wulzer}. 

Note that small values of $s_\theta$ require tuning, cf. \eqref{vacuum}.  
The largest irreducible tuning of $s_\theta $ is due to $\lambda_{UD}$, and can be quantified as
$|d \log s_\theta  /d \log \lambda_{UD} | = 2 \cot^2 \theta $. The tuning due to $f$ is, in principle, $50\%$ larger.  However, 
this is significantly reduced if $f$ and $m_H$ are correlated, e.g. via SUSY breaking (thus accounting for their proximity, see Table \ref{tab:examples}
and concluding remarks).

The light Higgs $h_1 VV$ ($V= W^\pm , Z $) and $h_1 \bar f f$ couplings normalized to the SM ones, $\kappa_V$ and $\kappa_F$, and their $s_\theta^2 \ll 1$ limits are
\beq
\begin{split}
 \kappa_V  
&=c_\alpha s_\beta -s_\alpha\,c_\beta\, c_\theta  ~\to ~1- c_\beta^2 \,s_\theta^2\, /2,\cr
\kappa_F &= c_\alpha / s_\beta ~ \to ~1-  c_{2\beta} \, c_\beta^2\, s_\theta^2\, /2  \,;
\label{higgspheno}\end{split}\eeq
constraints on these couplings from the LHC are at the level of 15\% and 25\% respectively \cite{LHChiggscouplings}.

There are two $Sp(4)$ covariant gauge field strengths \cite{Galloway:2010bp},
\beq 
{\cal D}_{\mu\nu}  = \nabla_{[\mu }d_{\nu ]} ,~~~  
{\cal F}_{\mu\nu}  = - i [\nabla_\mu  , \nabla_\nu]\,. 
  \label{FandD}
\eeq
$\nabla_\mu {\cal O}=  \partial_\mu {\cal O} - i [E_\mu ,{\cal O}]$  
is the $Sp(4)$ covariant derivative.
They transform homogeneously under $Sp(4)$, 
with ${\cal D}_{\mu\nu} $ a
5-plet and ${\cal F}_{\mu\nu}$ a 10-plet.
The effective operator
\beq {\cal L}_{\chi {\cal F}{\cal F}} = {\lambda_{\chi} \sec\beta  \sin \theta  \over 64 \pi^3 v_W } \, {\rm tr}(\chi^+ \,{\cal F}_{\mu\nu} \,{\cal F}^{\mu\nu}  )\,\label{chiFF} \eeq
($\lambda_\chi = \mathcal O(1)$ in NDA) induces an $h_1 \gamma\gamma$ coupling 
\beq {\cal L} = c^{\rm TC}_\gamma {\alpha \over \pi v_W} \,h_1 \,A_{\mu\nu} A^{\mu\nu},~~
c^{\rm TC} _\gamma    = {\lambda_{\chi} \, \lambda_{UD}\, c_\alpha\over 32 \sqrt{2 } \pi c_\beta} \,s_\theta^2\,,
\label{cgaga}\eeq
compared to $c_\gamma^{\rm SM} \simeq .23$. 
Including the modified Higgs couplings to $t, W$ in the $h_1 \to \gamma \gamma$ decay rate \cite{Carmi:2012in}, we obtain
\beq \Gamma_{\gamma\gamma} / \Gamma^{SM}_{\gamma\gamma}\simeq 1.52  \, |\kappa_F c_\gamma^{\rm SM} -1.04 \kappa_V +
c^{\rm TC} _\gamma |^2  \,.\label{gammagamma}\eeq
Thus, the TC shift in $\Gamma_{\gamma \gamma}$ is suppressed by $s_\theta^2$, like $\Gamma_{VV, \bar f f}$.

Significant effects enter beyond $\mathcal O(p^2)$, away from the chiral limit, e.g. in examples with $\hat m \sim f$.
However, our conclusions are not qualitatively altered \cite{GKMlong}: 
$t_\theta = -m_{UD}/m_{12}$ holds to all orders, as already argued; 
$v$, $\sin\theta$ and $m_h^2$ 
retain the forms given in \eqref{vacuum} 
and \eqref{mhp2}, up to negligible corrections from cubic and higher order Higgs couplings,
with $Z_2  \to Z_2 [1+  \mathcal O(\hat m/ 2 \pi f) ]$ and 
$m_H^2 \to m_H^2 + \mathcal O(\lambda_{U,D}^2 f^2 )$. 
As at $\mathcal O(p^2)$, $m_{h_1}^2$ is suppressed by $s_\theta^2$.
In \eqref{vWreln} and \eqref{tbeta}, $f$ is replaced by the full TC-pion decay constant,
$f \to f [1  + \mathcal O(\hat m / 4 \pi )]$.
Isospin and $G_{LR}$ combined imply that $\tilde \pi^3 - \pi^5$ mass mixing would be $\propto  \delta m_{12} \,\delta\lambda_{UD} \times  ( f  \sin\theta , v )$, thus first entering at $\mathcal O(\chi^{\pm\,2})$, or $\mathcal O(p^4)$.

There are two $Sp(4)$ singlet $C$-even scalar resonances of note, with masses of $\mathcal O(\Lambda)$: 
the $P$-even $\sigma $ and 
$P$-odd $\eta'$.  The $\sigma$ is not broad if $\sigma \to \tilde \pi \tilde \pi , h_2 h_2 $ are kinematically forbidden;
the $\eta'$ has a gluonic component due to the TC axial $U(1)$ anomaly.  
They have well defined $C$ and $P$ transformations, possessing dimension-5
$ \sigma AA$ and anomaly induced $\eta'  A\tilde{A}$ couplings, unlike the $\pi^{a}$ (however, anomalous $\pi^{5} (W\tilde W,Z\tilde Z,A\tilde Z)  $ couplings exist \cite{Gripaios:2009pe,Arbey:2015exa}).
The $\sigma$ induces the NDA shifts $\delta m_h^2 \sim - \lambda_{UD}^2 f^2 s_\theta^2 $; $\delta \kappa_V \sim \lambda_{UD}  c_\beta s^2_\theta  /(4\pi ) $; negligible 
$\delta \kappa_F $; and  $\delta c_\gamma^{\rm TC}/c_\gamma^{\rm TC} \sim 1$
\cite{GKMlong}.
\section{The vector resonances}
All resonances appear in representations of $Sp(4)$. 
We consider the lowest lying 10-
and 5-plet vectors (we do not consider the singlet here), 
\beq \hat R_{10} = R_{10}^a T^a,~~~\hat R_{5} = R_{5}^a X^a\,,\label{R510}\eeq
with $\hat R  \mapsto V \hat R V^\dagger$ under $Sp(4)$ (see also Ref.~\cite{Franzosi:2016aoo}).
Under $SU(2)_1\times SU(2)_2$, the $10 = (3,1) + (1,3) + (2,2)$,
where $R_{10}^{a\pm} =(R_{10}^{a} \pm R_{10}^{a+3})/\sqrt{2}$, $a$=1,2,3, are the two triplets.
$R^{1,2,3}_{10} $,$\,R^{4,5,6}_{10} $, and $\,{R}_5^{1..3}$ are triplets under $SU(2)_V$.
$G_{LR}$ interchanges $SU(2)_{1\leftrightarrow 2} $, 
in addition to $SU(2)_{L\leftrightarrow R}$, and $R_{10}^{a+} \leftrightarrow R_{10}^{a-}$.
The transformations of $\Psi^\dagger\bar\sigma_\mu T^a \Psi  $ imply $G_{LR}$-even (odd)
$R_{10}^{1,2,3}$ ($R_{10}^{4,5,6}$); and $CP$-even (odd) $R_{10}^{2,5}$ ($R_{10}^{1,3,4,6}$). $R^a_5$ transforms like $\pi^a$.
The Lorentz vector indices 
are also raised/lowered under $CP$.
Based on the vector currents for $R_{10}^{1,2,3}$ and $R_{5}^{1,2,3}$   
at $\theta=\pi/2$, $\hat R_{10}$ and $\hat R_5$ generalize
the 
QCD $\vec \rho$ and $\vec a_1$ triplets, respectively. 
However, $R_{10}^{a}$ and $R_{10}^{a+3}$, $a$=1,2,3, 
are the $G_{LR}$ ``parity doubling partners".

We use the antisymmetric tensor formalism for vectors 
\cite{Gasser:1983yg,Ecker:1988te,Ecker:1989yg}.  It is convenient for describing vector interactions with 
electroweak gauge fields, and avoids field redefinitions.
The kinetic terms are 
\beq {\cal L}_{\rm kin} =  -\frac12 \,{\rm tr}( \nabla^\lambda \hat R_{\lambda \mu} \nabla_\nu \hat R^{\nu \mu}  -\frac12 M_R^2 \hat R_{\mu\nu} \hat R^{\mu\nu} )\,,
\label{vectorkinetic}
\eeq
where 
$M_R^2$ is the mass in the chiral limit, 
and $R$ denotes $R_{5}$ or $R_{10}$. 
A related object,
\beq R_\mu = -M_R^{-1} \, \nabla^\nu  R_{\nu\mu}, \label{Rmu}\eeq
satisfies the massive Proca equation, 
and 
$\langle 0 | R_\mu | R \rangle = \epsilon_\mu $.

The most general $\mathcal O(p^2)$ Lagrangian, linear in $\hat R_{5,10}$, %,
\beq
\!\!{\cal L}^{(2)}_{R}\! ={\rm tr}\bigg(\hat {R}_{10,\mu\nu}\bigg[ {F_{10}\over \sqrt{2}} {\cal F}^{\mu\nu}\!+{i}\, G_{10}  d^\mu d^\nu \bigg]\! +  \!
{F_{5}\over \sqrt{2}}\hat {R}_{5,\mu\nu}{\cal D}^{\mu\nu}\bigg) \label{L2R}
\eeq
yields the bilinears ($a$=1,2,3)
\bea \begin{split}
 {\cal L}&_{\rm bilinear} =-{1\over 4}{F_{10}}\,
{R}_{10}^a  \big(g_2 W_{}^a + g_1 B_{} \,\delta^{a3} \big) \label{bilinears} \\
&-\frac14 \big({F_{10}} \,c_\theta\,  {R}_{10}^{a+3}  -{ F_{5 }}\,s_\theta \, {R}_{5}^a \big) \big(g_2 W_{}^a  - g_1 B_{} \,\delta^{a3} \big),  
\end{split}
\eea
where $F_{10,5}$ are the vector decay constants, 
\beq  \langle {R}_{10 (5)}^a | \Psi^\dagger \, \bar \sigma_\mu\,T^a (X^a) \,\Psi |0\rangle = -i F_{10 (5)} M_{10 (5)}\epsilon_\mu^*  \,,\label{decayconstants}\eeq
with $M_{10,5}$ the total masses.
They induce $R_{5,10}$ couplings to the SM fermions, responsible for vector Drell-Yan (DY) production, and obtained via the following substitutions in the SM couplings,
\beq 
\begin{split}
W_\mu^a  &\to W_\mu^a - {g_2 F_{10}\over 2 M_{10 }} \left(R_{10,\mu}^a + R_{10,\mu}^{a+3}c_\theta \right) +   {g_2 F_{5}\over 2 M_{5} } R_{5,\mu}^a s_\theta ~~\cr
B_\mu & \to B_\mu - {g_1F_{10}\over 2 M_{10} } \left(R_{10,\mu}^3 - R_{10,\mu}^{6} c_\theta \right) -   {g_1F_{5}\over 2 M_{5} }  R_{5,\mu}^3 s_\theta\,.
\end{split}
\label{Wsub}
\eeq
The leading $R_{10}$ decays originate from the ${\cal L}^{(2)}_R$ trilinears,%
\beq 
\begin{split}
-{ G_{10}  M_{10} \over 2\sqrt{2}  f^2} &\big(\,\epsilon^{abc} {R}_{10,\mu}^{a} \pi^b \partial^\mu \pi^c \cr &~~~+{R}^{a+3}_{10,\mu}  \big[\pi^5  \partial^\mu \pi^{a} - \pi^{a} \partial^\mu \pi^5 \big]\,\big) + \dots , 
\end{split}\label{Gpipi}
\eeq
where the ellipses denote couplings of $R_{10}^{7,\dots,10}$. 
In the vector meson dominance (VMD) approximation, $G_{10 } = -2 \sqrt{2} f^2/F_{10} $  (the VMD $\rho\pi\pi $ coupling, $g_{\rho\pi\pi} =- m_\rho / f_\rho$, is $16\%$ below experiment; $\phi KK$ is within a few $\%$).
Projecting \eqref{Gpipi} onto the $G^a$ gives couplings to longitudinal $W_L , Z_L$.
For $\hat m \sim f$, $R^{1,2,3, (4,5,6)}_{10} \!\to \!\tilde \pi \tilde \pi, (h_2 \tilde \pi)$ are closed, and $R^{1,2,3 (4,5,6)}_{10} \!\to\! \tilde \pi \,W_L /Z_L, (\tilde \pi h_1, h_2\, W_L /Z_L )$ dominate.

The UV contribution to the $S$ parameter can be parametrized in terms of the effective 
Lagrangian coupling, $-g_1 g_2  S_{UV}/(32 \pi ) \,   W^3_{\mu\nu} B^{\mu\nu}  $.   
Tree-level $R_{10}^{3,6}$, $R_{5}^3$ exchange, cf. \eqref{bilinears}, thus yields 
\beq \Delta S_{\rm tree} = 4\pi \left({F_{10}^2 / M^2_{10} } - {F_{5}^2 / M^2_{5} } \right) \sin^2\theta. \label{SUV}\eeq
The $s_\theta^2$ suppression is a feature of misalignment \cite{Agashe:2004rs,Barbieri:2007bh,contino,Marzocca:2012zn,wulzer} ($S_{UV}$ is $\Delta I=1$, and the underlying operators 
${\rm tr} ( 2{\cal F}^2)$, $ {\rm tr}  ({\cal D}^2)\supset \pm \,g_1 g_2 W^3_{\mu\nu} B^{\mu\nu}  s_\theta^2 /2 $  \cite{Galloway:2010bp}).  It has an explicit origin in \eqref{SUV}: $R_{10}^{3,6}$ 
parity doubling cancelation $\propto 1-c^2_\theta$; and $s_\theta$ suppression of the $R^3_5$ couplings. 
The scalar loops in $S$ are log divergent, 
due to a $c_\theta $ factor in the $\pi^4$ gauge boson couplings.
After subtracting the SM Higgs, 
\beq \begin{split} 
\Delta S_{\rm loop}^{}= {1\over 24 \pi} \left(s_\theta^2\,\log {\Lambda^2\over m_{h_1}^{2 s_\alpha^2} m_{h_2}^{2 c_\alpha^2 } }+ F_{\rm fin}\right)\!,  \end{split}\label{SIRfin}
\eeq
where $F_{\rm fin}$ contains finite loop contributions \cite{GKMlong}. The first term receives a $s_\alpha^2$ suppression that is not present in the composite Higgs case, due to projection of $\pi^4$ onto $h$. For cut-off $\Lambda \le 8 \pi f$,
we find $\Delta S_{\rm loop}^{} < 0.01$ in our examples.  A more refined dispersion integral approach containing higher order $R_{5,10}$ contributions \cite{rychkov} would eliminate the divergence, with $S$ remaining a small effect.

The $T$ parameter arises from: (i) scalar loops with isospin breaking entering via 
$\tilde \pi^3\!-\!\eta',\pi^5$ mixings,  and $\tilde \pi^3 \!-$$ \tilde \pi^+$ mass splitting; 
(ii) $G^+ $ wave function renormalization via $B\!-\!R_{5,10}^{1,2,4,5} $ loops. 
The loops in (i) vanish in the $\delta \lambda_{UD} \to 0 $ limit, and in (ii) they are $c_\beta^2$ suppressed (due to
projection 
onto $G^+ $) compared to 
the composite Higgs and TC analogs \cite{rychkov, kamenik,pich}.
Thus, $S$ and $T$ reasonably lie within the allowed $1\sigma$ ellipse \cite{STellipse,PDG}.

Fermion mass corrections to $M_{10, 5}$ arise from 
terms $\propto {\rm tr}( \hat R_{10,5}^2 \chi_+ ) $ at $\mathcal O(p^2)$, and larger $\chi^+$ multiplicities at higher orders.
In the limit $\delta m_{12}, \delta m_{UD} \to 0$ they respect $Sp(4)$, and are $\theta$-independent polynomials in $\hat m$, 
as seen from \eqref{MlambdaSO5}. This is true of all corrections to the chiral limit.
Thus, the $\theta=\pi/2$ lattice measurements \cite{Arthur:2016dir} (also see \cite{Detmold:2014kba}) of $f_\pi$ (full decay constant), $m^2_\pi$, and $M_{10,5}$  hold for arbitrary $\theta$. 
$F_{10}$ can be estimated
by scaling a fit to the quark mass dependence of QCD vector decay constants, normalized to the observed pion decay constant.  Using $f_{\rho , \omega,\phi}$, and 
the lattice heavyonium decay constant between $m_{J/\Psi}$ and $m_\Upsilon$ \cite{McNeile:2012qf, HPQCD}, yields
a function ${\cal F}$ such that $F_{10}/f  \approx {\cal F}[ \hat m /f  ] \, (f_\rho\,/ f_\pi )_{\rm QCD} $ (or $F_{10}/f \approx [1.6,1.8]$ for 
$\hat m /f = [0,1.5]$) 
\cite{kagantalks,GKMlong}.
The contribution of $R_5$ in \eqref{SUV} is bounded via the approximate upper and lower bounds, $M_5<  M_{10}  \,m_{a_1} /m_\rho $ 
($M_{5}/M_{10}$ decreases beyond the chiral limit, 
approximated by the 
QCD ratio) and $F_5 > f_{a_1}  f / f_\pi^{\rm qcd}  $ 
($F_5$ increases away from the chiral limit, obtained by scaling from QCD).
We take $f_{a_1}=152$ MeV  \cite{Isgur:1988vm} for the poorly known decay constant, using an updated 
${\rm Br}(\tau^+ \to \nu_\tau \pi^+ \pi^+ \pi^- )$.

The above procedure yields $\Delta S_{\rm tree} / s_\theta^2 <   [0.11,0.09]$  ([0.19, 0.13] for $R_{10}$) 
for $\hat m /f  = [0 , 1.5]$, confirming that agreement with the observed $1\sigma $ range
$\Delta S=  0.10 \pm 0.08$ \cite{STellipse}, $0.00\pm 0.08$ \cite{PDG} is reasonable.   
The significant decrease in the $R_{10}$ contribution  
away from the the chiral limit reflects the greater vector mass
vs. decay constant quark mass dependence observed in QCD.

\section{Examples}
The vector masses 
can span a  
wide range, due to the freedom to vary $\sin\theta$ and the TC fermion masses. 
This is illustrated in the representative examples of Table \ref{tab:examples}, for different values of the UV inputs $f,m_H ,m_{12},\lambda_{UD}$ (with $\delta m_{12},\,\delta \lambda_{UD}=0$), where the two isotriplet charged vectors are defined as $r_{1(2)}^\pm \equiv (R_{10}^{1(4)} \mp i R_{10}^{2(5)})/\sqrt{2}$.
The vacuum alignment and scalar spectrum have been obtained at $\mathcal O(p^2)$, with $Z_2 = 1.47$ \cite{Arthur:2016dir}, imposing the $m_{h_1} = 125$ GeV, $v_W = 246$ GeV, and neglecting the Higgs quartic for simplicity. 
The tuning of $s_\theta =.6,.4,.3$ (due to $\lambda_{UD}$) is approximately $30\%$, $10\%$, $5\%$, respectively.
Note that $m_H$ is essentially fixed by $\sin\theta$; both $m_H$ and $f$ (or $\Lambda_{\rm TC} \sim 4 \pi f $) increase as $\sin\theta$ decreases; while for given $\sin \theta$,  $f$ increases as $m_{12}$ and $\hat m$
decrease. In all of the examples, the deviations in $\kappa_V$ and $\kappa_F$ from 1 (SM) are $< 1\%$, and the deviations in 
the Higgs diphoton decay width from its SM value are $< 2\% $ for $|\lambda_\chi |\le  2$, cf. \eqref{cgaga},\eqref{gammagamma} (with the exception of the first example, where a deviation as large as 6\% is possible).

$M_{10}$ follows from \cite{Arthur:2016dir} and $F_{10}$ from the scaling described above.
The vector decay widths follow from the VMD estimate for $G_{10}$.  The narrow width approximation is used throughout.
In the first example, the $R_{10} \to \tilde \pi\tilde \pi , h_2 \tilde \pi$ channels are closed, yielding relatively narrow widths.
In the last two examples, with $\hat m/f \sim 0.1$, the phase space suppression in the $R_{10} \to \tilde \pi\tilde \pi ,h_2 \tilde \pi$ channels is small, 
yielding very large widths, thus the narrow width approximation is rough.
%
%in fact beyond the validity of the narrow width approximation is 
%yielding large widths and negligible $W,Z,h$ diboson branchings.
In general, the combination of  $F_{10} /M_{10} \sim 0.1$ and small branching fractions to pairs of gauge bosons implies that the vector resonances are safe (by at least $\mathcal O(10-100) $ for $M_{10} \sim 1 - 3$ TeV) from current LHC bounds.
\begin{table*}[ht]
\begin{center}
\renewcommand{\arraystretch}{1.1}
\begin{tabular}{|c | c | c | c || c| c | c|c|c|c|c||c|c|c|c|c|c|c|c|}
\hline
$f$ &$m_{12}$ &$m_H$ & $\lambda_{UD}$ & $s_\theta$ &  $c_\beta $ & $s_\alpha$  & $v $ &$ \hat m $ & $m_{\tilde \pi }$ & $m_{h_2}$ &   $M_{10}$ & $F_{10}$ &  $\sigma_{pp\to r_1^+ }$ [fb] 
&$\sigma_{pp\to r_2^+ }$ [fb]  & $ {\rm Br}_{r^+_1  \to WZ} $ & $ {\rm Br}_{r^+_2 \to h_1 W}$ & $\Gamma_{r_1^+} $& $\Gamma_{r_2^+}$ \\
\hline
93 &190 &212&0.84& 0.6 &   0.23 & 0.19 & 240 &119 & 930& 922 & 1585 & 167 &5.28 &3.38 &0.089 &0.072 &29      &  23    \\
165 &200 &323& 0.52& 0.4 &   0.27 & 0.25 & 237 & 109 & 1199 & 1193  & 2504 & 282&0.50& 0.42& 0.06 & 0.06 & 115  & 106   \\
232 &240 &433& 0.45& 0.3 &   0.28 & 0.27 & 236 &126 &1531 & 1526 &3429 & 392& 0.054 & 0.049 &0.04 & 0.04 & 279   &  271  \\
\hline
261  &50 &341&0.14& 0.4 &   0.43 & 0.41 & 223 &27 & 802& 792 &3538 & 427 &0.046&0.039 &0.043&0.039 &1228   & 1219\\
367  &60 &463&0.12 & 0.3 &   0.45 & 0.44 & 220 &31 &1033 & 1025 &4950 & 599 &$0.001$&$0.001$ &0.05&0.05 &1799    &1793\\
\hline
\end{tabular}   
%\linespread{1}
\caption{\small{Examples of vacuum alignment, scalar spectrum, and $R_{10}$ properties, 
see text (all masses are in GeV).}}
\label{tab:examples}
\end{center}
\end{table*} 

\section{Discussion}
BTC with $N_c = n_f=2$ provides the minimal UV complete realization of the partially composite-pNGB Higgs. Several other noteworthy features 
are summarized below:
(i) $\Lambda_{TC}\gsim $3 TeV and an enhanced Higgs mass parameter, e.g. $m_H \sim 3\, m_h$, are 
accesible with moderate $s_\theta$,  
or tuning.  
(In the TC-vacuum, sub-10\% deviations in the Higgs couplings would require  $f<100$ GeV, or $\Lambda_{\rm TC} \lsim 1$ TeV \cite{Chang:2014ida,kagantalks});
(ii) deviations from the SM Higgs couplings of  $\mathcal O(1\%)$ are typical, due to suppression by 
$s_\theta^2\,c_\beta^2 $; 
(iii) 
The ratio $f / M_{10}$ on the lattice
suggests that agreement with the $S$ parameter at $1 \sigma$ is realized at moderate $s_\theta$;
while potentially dangerous 
$T$ parameter loops are $c_\beta^2 = \mathcal O(0.1)$ suppressed; 
(iv) detection of vector mesons  
at the LHC will be challenging. 

Our ultimate goal is natural EW symmetry breaking.  
In the present context this would involve linking the size of the Higgs mass parameter to the TC scale, $\Lambda_{\rm TC} = \mathcal O(3\, {\rm TeV}$).
One direction that we are exploring is embedding our setup into a supersymmetric theory with Dirac gauginos.  Supersymmetrized minimal BTC is formulated as supersymmetric QCD with $N_c \! \! =\! n_f \! \! = \! \!2$ and one adjoint matter superfield.  This theory is known to have a strong IR fixed point, with unbroken chiral symmetry \cite{Elitzur:1995xp,Elitzur:1995sx}; a two-loop estimate yields a fixed point coupling $\alpha^* \approx 1.8$. 
A direct link between the scale of TC superpartner masses and $\Lambda_{\rm TC}$ can be realized if the Dirac TC-gaugino and scalar matter fields decouple in the superconformal region
\cite{Azatov:2011ht,Azatov:2011ps}.\footnote{Majorana gaugino masses would undergo power law running $m_{\tilde g} (\mu) \propto (\mu/\Lambda )^{\gamma'}$ with $\gamma'  >0$ in the superconformal region \cite{Luty:1999qc}. Remarkably, we find $\gamma'  <0 $ in the Dirac case, with moderate $\gamma' \approx -0.41$ at two loops, rendering the solution to the scale coincidence problem a viable one.}  
Integrating out these massive states triggers a confining phase with $\Lambda_{\rm TC} \lesssim m_{\tilde g_{\rm TC}}$.  
The Higgs mass, $m_H \sim m_{\tilde W}/4\pi$ generated with finite loops of Dirac EW gauginos \cite{Fox:2002bu,Alves:2015kia}, can then naturally be of order $f \sim \Lambda_{\rm TC}/4\pi$ if the SM gaugino masses satisfy $m_{\tilde W} \sim m_{\tilde g_{\rm QCD}} \lesssim m_{\tilde g_{\rm TC}}$.
Furthermore, the resulting effective theory contains 4-technifermion operators which affect the vacuum alignment and may allow a construction without the explicit singlet masses $m_{1,2}$.

Further study of potential UV completions, as well as detailed collider phenomenology will be presented elsewhere.\\

%%%%%%%%%%%%%%%%%%%%%%%%%%%%%%%%
\mysection{Acknowledgements}
%%%%%%%%%%%%%%%%%%%%%%%%%%%%%%%%
A.K. thanks Claudio Pica, Francesco Sannino, and William Detmold for discussions of lattice results, and J.G. thanks Duccio Pappadopulo for many helpful conversations.  
 A.K. and A.M thank Seung Lee, Patipan Uttayarat, and Jure Zupan for collaboration on related work.  
 A.K. thanks the MITP for their hospitality and Advanced Grant EFT4LHC of the European Research Council (ERC) for support in the closing stages of this work.
The work of A.K. is supported by the DOE grant DE-SC0011784. A. M. was partially supported by the National Science Foundation under Grant No. PHY-1417118.  

\end{document}